\documentclass[12pt,prd,tightenlines,nofootinbib,showpacs]{revtex4}
\usepackage{bm}
\usepackage{graphics}
\usepackage{epsfig}
\begin{document}
\title{\begin{flushright}{\rm\normalsize SSU-HEP-10/7\\[5mm]}\end{flushright}
HYPERFINE STRUCTURE OF THE GROUND \\STATE MUONIC ${\rm ^3He}$ ATOM}
\author{A.A.Krutov, A.P.Martynenko}
\affiliation{Samara State University, Pavlov street 1, 443011,
Samara, Russia}

\begin{abstract}
On the basis of the perturbation theory in the fine structure
constant $\alpha$ and the ratio of the electron to muon masses we
calculate one-loop vacuum polarization and electron vertex
corrections and the nuclear structure corrections to the hyperfine
splitting of the ground state of muonic helium atom $(\mu\ e \
^3_2He)$. We obtain total result for the ground state hyperfine
splitting $\Delta \nu^{hfs}=4166.471$ MHz which improves the
previous calculation of Lakdawala and Mohr due to the account of new
corrections of orders $\alpha^5$ and $\alpha^6$. The remaining
difference between our theoretical result and experimental value of
the hyperfine splitting lies in the range of theoretical and
experimental errors and requires the subsequent investigation of
higher order corrections.
\end{abstract}


\maketitle

\section{Introduction}

Muonic helium atom $(\mu\ e \ ^3_2He)$ represents the simple
three-body atomic system consisting of one electron, negative
charged muon and positive charged helion ($^3_2He$). Contrary to the
muonic helium atom $(\mu\ e \ ^4_2He)$ it has more complicated
ground state hyperfine structure which appears due to the
interaction of the magnetic moments of the electron, muon and helion
\cite{LM3,HH1,HH2,RD1,Chen1,Chen2,Amusia1,Krivec,AF}. The
investigation of the energy spectrum of this three-particle bound
state is important for the further check of quantum electrodynamics.
Moreover, light muonic atoms (muonic hydrogen, muonic helium, ions
of muonic helium etc.) represent a unique laboratory for precise
determination of the nuclear properties such as the nuclear charge
radius \cite{Nature}. Hyperfine splitting (HFS) of the ground state
of muonic helium atom $(\mu\ e \ ^3_2He)$ was measured many years
ago with sufficiently high accuracy \cite{Gladish}:
\begin{equation}
\Delta\nu^{hfs}_{exp}=4166.3(2)~MHz.
\end{equation}
There are two approaches to the calculation of the energy spectrum
of muonic helium atom $(\mu\ e \ ^3_2He)$. First approach in
\cite{LM3,HH1,Amusia1} is based on the perturbation theory (PT) for
the Schr\"odinger equation. In this case there is the analytical
solution for the three particle bound state wave function in the
initial approximation. Using it the analytical calculation of
different corrections to HFS can be performed. Contrary to the
energy levels of two-particle bound states which were accurately
calculated in quantum electrodynamics
\cite{HBES,EGS,SGK,M2004,M3,M4}, the hyperfine splitting of the
ground state in muonic helium atom was calculated on the basis of
the perturbation theory with essentially less accuracy. Another
approach is built on the variatinal method
\cite{RD1,Krivec,AF,Chen1,Chen2,AF1,VK2000,KP2001} which allows to
increase the accuracy of the calculation. In the beginning, the
accuracy of the HFS calculation was not sufficiently high because
corrections of six order in $\alpha$ were estimated only
approximately. A feature that distinguishes light muonic atoms among
the simplest atoms is that the structure of their energy levels
strongly depends on the vacuum polarization, nuclear structure and
recoil effects. Subsequently, the corrections of order $\alpha^2$ to
hyperfine splitting were studied in \cite{Chen1,Chen2} on the basis
of variational and global-operator method. The theoretical value of
HFS obtained in \cite{Chen1,Chen2} contains very small uncertainty
and agrees with the experimental value (1).

In this work, which continues our investigation \cite{KM2008}, we
aim to refine the calculation of Lakdawala and Mohr \cite{LM3} using
their analytical approach to the description of the muonic helium
atom. We investigate such contributions of the one-loop electron
vacuum polarization of order $\alpha^5M_e/M_\mu$ and the nuclear
structure of orders $\alpha^5, \alpha^6$ which are significant for
the improvement of the theoretical value of the hyperfine splitting
obtained in \cite{LM3} on the basis of perturbation theory. Another
purpose of our study consists in the improved calculation of the
electron one-loop vertex corrections to HFS of order $\alpha^5$
using the analytical expressions of the Dirac and Pauli form factors
of the electron.

The bound particles in muonic helium atom have different masses
$m_e\ll m_\mu\ll m_\alpha$. As a result the muon and helion compose
the pseudonucleus $(\mu\ ^3_2He)^+$ and the muonic helium atom looks
as a two-particle system in the first approximation. Three-particle
bound system $(\mu\ e \ ^3_2He)$ is described by the Hamiltonian
\cite{LM1,LM2}:
\begin{equation}
H=H_0+\Delta H+\Delta H_{rec},~~~H_0=-\frac{1}{2M_\mu}\nabla^2_\mu-\frac{1}{2M_e}
\nabla^2_e-\frac{2\alpha}{x_\mu}-\frac{\alpha}{x_e},
\end{equation}
\begin{equation}
\Delta H=\frac{\alpha}{x_{\mu e}}-\frac{\alpha}{x_e},~~~\Delta
H_{rec}=-\frac{1}{m_h}
{\mathstrut\bm\nabla}_\mu\cdot{\mathstrut\bm\nabla}_e,
\end{equation}
where ${\bf x_\mu}$ and ${\bf x_e}$ are the coordinates of the muon
and electron relative to the helium nucleus, $M_e=m_em_h/(m_e+m_h)$,
$M_\mu=m_\mu m_h/(m_\mu+m_h)$ are the reduced masses of subsystems
$(e\ ^3_2He)^+$ and $(\mu\ ^3_2He)^+$. In the initial approximation
the wave function of the ground state has the form
\cite{LM1,LM2,Borie}:
\begin{equation}
\Psi_0({\bf x_e},{\bf x_\mu})=\psi_e({\bf x_e})\psi_\mu({\bf
x_\mu})=\frac{1}{\pi} (2\alpha^2M_eM_\mu)^{3/2}e^{-2\alpha M_\mu
x_\mu}e^{-\alpha M_e x_e}.
\end{equation}

The hyperfine interaction in the ground state in $(\mu\ e\ ^3_2He)$
is determined by the following Hamiltonian:
\begin{equation}
\delta H=-\frac{8\pi}{3}({\bm\mu}_N\cdot{\bm\mu}_\mu)\delta({\bf
x}_\mu)-\frac{8\pi}{3}({\bm\mu}_\mu\cdot{\bm\mu}_e)\delta({\bf
x}_\mu-{\bf
x}_e)-\frac{8\pi}{3}({\bm\mu}_e\cdot{\bm\mu}_N)\delta({\bf x}_e),
\end{equation}
where ${\bm\mu}_e=-g_ee/(2m_e){\bf s}_e$, ${\bm\mu}_\mu=-g_\mu
e/(2m_\mu){\bf s}_\mu$, ${\bm\mu}_N=-g_Ne/(2m_p){\bf I}_N$ are
magnetic moments of the electron, muon and helion. The total spin of
the three spin-1/2 particles can be either 3/2 and 1/2. The matrix
element of (5) leads to the shift of the energy levels which takes
on form:
\begin{equation}
\delta E=<\delta H>=-a\ ({\bf I}_N\cdot{\bf s}_\mu)-b\ ({\bf
s}_\mu\cdot{\bf s}_e)-c\ ({\bf s}_e\cdot{\bf I}_N),
\end{equation}
where
\begin{equation}
a=\frac{2\pi\alpha}{3}\frac{g_Ng_\mu}{m_pm_\mu}<\delta({\bf
x}_\mu)>,~b=\frac{2\pi\alpha}{3}\frac{g_\mu g_e}{m_\mu
m_e}<\delta({\bf x}_\mu-{\bf
x}_e)>,~c=\frac{2\pi\alpha}{3}\frac{g_eg_N}{m_em_p}<\delta({\bf
x}_e)>.
\end{equation}

The diagonalization of the matrix element $<\delta H>$ gives three
eigenvalues:
\begin{equation}
\nu_{1,2}=\frac{1}{4}(a+b+c)\pm\frac{1}{2}(a^2+b^2+c^2-ab-bc-ca)^{1/2},~
\nu_3=-\frac{1}{4}(a+b+c).
\end{equation}
The values $\nu_{1,2}$ and $\nu_3$ correspond to the total angular
momentum $\frac{1}{2}$ and  $\frac{3}{2}$. In the case of muonic
helium $(\mu\ e \ ^3_2He)$ we have the relations $a\gg b$ and $a\gg
c$. So, the eigenvalues $\nu_{1,2}$ can be written with good
accuracy in a more simple form:
\begin{equation}
\nu_1=\frac{3}{4}a+\ldots ,
~~~\nu_2=-\frac{1}{4}a+\frac{1}{2}(b+c)+\ldots .
\end{equation}
As a result the smaller hyperfine splitting interval related to the
experiment (1) is given by
\begin{equation}
\Delta\nu^{hfs}=\nu_2-\nu_3=\frac{3}{4}(b+c)+O(\frac{b}{a},\frac{c}{a}).
\end{equation}

Basic contributions to the coefficients a, b, c were calculated
analytically in \cite{LM3} from the contact interaction (5) in the
first and second order PT. Taking into account numerical values of
gyromagnetic factors $g_e=2$ for the $b$ coefficient,
$g_e=2(1+\kappa_e)=2(1+ 1.1596521859\cdot 10^{-3})$ for the $c$
coefficient, $g_\mu=2(1+\kappa_\mu)=2\cdot (1+1.16592069(60)\cdot
10^{-3})$, $g_N=2\cdot 2.127497718(25)$, we obtain for them:
\begin{equation}
b_0=\nu_F=\frac{8\alpha(\alpha
M_e)^3}{3m_em_\mu}=4516.307~MHz,~~~b_1=\kappa_\mu \nu_F=5.266~MHz,
\end{equation}
\begin{equation}
b_2=\nu_F(1+\kappa_\mu)\left(-3\frac{M_e}{M_\mu}+\frac{2}{3}S_{1/2}
\left(\frac{M_e}{M_\mu}\right)^{3/2}+\left(\frac{M_e}{M_\mu}\right)^2\ln\frac{M_\mu}{M_e}-
\frac{7}{64}\left(\frac{M_e}{M_\mu}\right)^2\right)=-64.322~MHz.
\end{equation}
\begin{equation}
c_0=\nu_F\frac{g_eg_N}{4}\frac{m_\mu}{m_p}=1083.256~MHz,
\end{equation}
\begin{equation}
c_1= c_0\left(\frac{3}{2}\frac{M_e}{M_\mu}+
\left(\frac{M_e}{M_\mu}\right)^2\ln\frac{M_\mu}{M_e}+(\ln
2+\frac{1}{4})\left(\frac{M_e}{M_\mu}\right)^2\right)=8.323~MHz.
\end{equation}
Note, that, as we determine contributions to the energy spectrum
numerically, the corresponding results are presented with an
accuracy of 0.001 MHz. We express further the hyperfine splitting
contributions in the frequency unit using the relation $\Delta
E^{hfs}=2\pi\hbar\Delta \nu^{hfs}$. Modern numerical values of
fundamental physical constants are taken from the paper \cite{MT}:
the electron mass $m_e=0.510998910(13)\cdot 10^{-3}~GeV$, the muon
mass $m_\mu=0.1056583668(38)~GeV$, the fine structure constant
$\alpha^{-1}=137.035999679(94)$, the helion mass $m(^3_2He)$ =
2.808391383(70)~GeV, the muon anomalous magnetic moment
$\kappa_\mu=1.16592069(60)\cdot 10^{-3}$, the electron anomalous
magnetic moment $\kappa_e= 1.15965218111(74)\cdot 10^{-3}$.

\begin{figure}[htbp]
\centering
\includegraphics{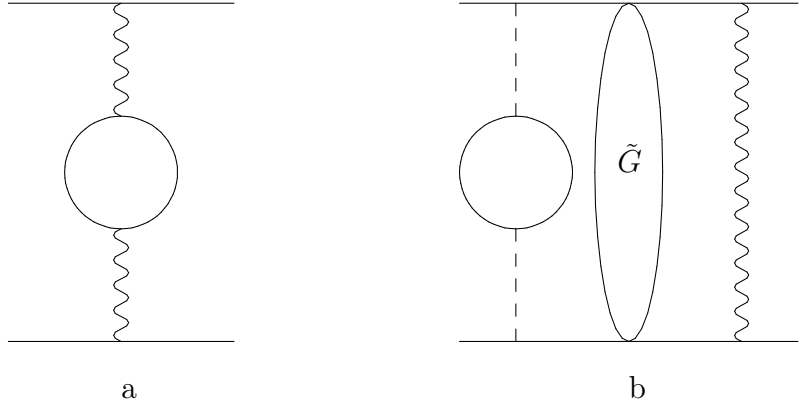}
\caption{The vacuum polarization effects. The dashed line represents
the Coulomb photon. The wave line represents the hyperfine part of
the Breit potential. $\tilde G$ is the reduced Coulomb Green's
function.}
\end{figure}

\section{Effects of the vacuum polarization}

The vacuum effects change the interaction (2), (3), (5) between
particles in muonic helium atom. One of the most important
contributions to HFS is determined by the one-loop vacuum
polarization (VP) and electron vertex operator. Indeed, the vacuum
loop leads to additional factor $\alpha/\pi$ in the interaction
operator, so that corresponding correction to HFS is of the fifth
order over fine structure constant. At the same time, the electron
vacuum polarization and vertex corrections to the hyperfine
splitting of the ground state contain the parameter equal to the
ratio of the Compton wave length of the electron and the radius of
the Bohr orbit in the subsystem $(\mu ^3_2He)^+$: $m_\mu\alpha/m_e$=
$1.50886\ldots$ . It appears in the matrix elements with the use of
the bound state wave function in which the characteristic momentum
is of order $m_\mu\alpha$. It is impossible to use the expansion
over $\alpha$ for such contributions to the energy spectrum. So, we
calculate them performing the analytical or numerical integration
over the particle coordinates and other parameters without an
expansion in $\alpha$. The effect of the electron vacuum
polarization leads to the appearance of a number of additional
corrections to the Coulomb potential which we present in the form
\cite{t4,EGS}:
\begin{equation}
\Delta V_{VP}^{eh}(x_e)=\frac{\alpha}{3\pi}\int_1^\infty
\rho(\xi)\left(-\frac{2\alpha}{x_e}\right) e^{-2m_e\xi
x_e}d\xi,~~~\rho(\xi)=\frac{\sqrt{\xi^2-1}(2\xi^2+1)}{\xi^4},
\end{equation}
\begin{equation}
\Delta V_{VP}^{\mu h}(x_\mu)=\frac{\alpha}{3\pi}\int_1^\infty
\rho(\xi)\left(-\frac{2\alpha}{x_\mu}\right) e^{-2m_e\xi x_\mu}d\xi,
\end{equation}
\begin{equation}
\Delta V_{VP}^{e\mu}(|{\bf x}_e-{\bf
x}_\mu|)=\frac{\alpha}{3\pi}\int_1^\infty
\rho(\xi)\frac{\alpha}{x_{e\mu}} e^{-2m_e\xi x_{e\mu}}d\xi,
\end{equation}
where $x_{e\mu}=|{\bf x}_e-{\bf x}_\mu|$. They give contributions to
the hyperfine splitting in the second order perturbation theory and
are discussed below. In the first order perturbation theory the
contribution of the vacuum polarization is connected with the
modification of the hyperfine splitting part of the Hamiltonian (5)
(the diagram (a) in Fig.1). In the coordinate representation it is
determined by the integral expression \cite{M1,M2,EM2009}:
\begin{equation}
\Delta V_{VP}^{hfs,e\mu}({\bf x}_{e\mu})=-\frac{8\alpha}{3m_em_\mu}
({\bf s}_e\cdot {\bf s}_\mu)\frac{\alpha}{3\pi}\int_1^\infty
\rho(\xi)d\xi\left[\pi\delta({\bf
x_{e\mu}})-\frac{m_e^2\xi^2}{x_{e\mu}}e^{-2m_e\xi x_{e\mu}} \right],
\end{equation}
\begin{equation}
\Delta V_{VP}^{hfs,eh}({\bf x}_e)=-\frac{8\alpha g_N}{6m_em_p} ({\bf
s}_e\cdot {\bf I}_N)\frac{\alpha}{3\pi}\int_1^\infty
\rho(\xi)d\xi\left[\pi\delta({\bf
x_e})-\frac{m_e^2\xi^2}{x_e}e^{-2m_e\xi x_e} \right].
\end{equation}

Averaging the potential (18) over the wave function (4) we obtain
the following contribution to the hyperfine splitting:
\begin{equation}
b_{\ VP}=\frac{8\alpha^2}{9m_em_\mu}\frac{(\alpha M_e)^3(2\alpha
M_\mu)^3} {\pi^3}\int_1^\infty\rho(\xi)d\xi\int d{\bf x}_e\int d{\bf
x}_\mu e^{-4\alpha M_\mu x_\mu}e^{-2\alpha M_ex_e}\times
\end{equation}
\begin{displaymath}
\times\left[\pi\delta({\bf x_\mu}-{\bf x}_e)-\frac{m_e^2\xi^2}{|{\bf x}_\mu-{\bf x}_e|}\right]
e^{-2m_e\xi|{\bf x}_\mu-{\bf x}_e|}.
\end{displaymath}
There are two integrals over the muon and electron coordinates in
Eq.(20) which can be calculated analytically:
\begin{equation}
I_1=\int d{\bf x}_e\int d{\bf x}_\mu e^{-4\alpha M_\mu x_\mu}e^{-2\alpha M_ex_e}\pi
\delta({\bf x_\mu}-{\bf x}_e)
=\frac{\pi^2}{8\alpha^3M_\mu^3\left(1+\frac{M_e}{2M_\mu}\right)^3},
\end{equation}
\begin{equation}
I_2=\int d{\bf x}_e\int d{\bf x}_\mu e^{-4\alpha M_\mu x_\mu}e^{-2\alpha M_ex_e}\frac{1}
{|{\bf x}_\mu-{\bf x}_e|}e^{-2m_e\xi|{\bf x}_\mu-{\bf x}_e|}=
\end{equation}
\begin{displaymath}
=\frac{32\pi^2}{(4\alpha
M_\mu)^5}\frac{\left[\frac{M_e^2}{4M_\mu^2}+\left(1+
\frac{m_e\xi}{2M_\mu\alpha}\right)^2+\frac{M_e}{2M_\mu}\left(3+\frac{m_e\xi}{M_\mu
\alpha}\right)\right]}{\left(1+\frac{M_e}{2M_\mu}\right)^3\left(1+\frac{m_e\xi}
{2M_\mu\alpha}\right)^2\left(\frac{M_e}{2M_\mu}+\frac{m_e\xi}{2M_\mu\alpha}\right)^2}.
\end{displaymath}
They are divergent separately in the subsequent integration over the
parameter $\xi$. But their sum is finite and can be written in the
integral form:
\begin{equation}
b_{\ VP}=\nu_F\frac{\alpha M_e}{6\pi
M_\mu\left(1+\frac{M_e}{2M_\mu}\right)^3}\int_1^\infty\rho(\xi)d\xi
\frac{\left[\frac{M_e}{2M_\mu}+2\frac{m_e\xi}{2M_\mu\alpha}\frac{M_e}{2M_\mu}+
\frac{m_e\xi}{2M_\mu\alpha}\left(2+\frac{m_e\xi}{2M_\mu
\alpha}\right)\right]}{\left(1+\frac{m_e\xi}
{2M_\mu\alpha}\right)^2\left(\frac{M_e}{2M_\mu}+\frac{m_e\xi}{2M_\mu\alpha}\right)^2}=0.036~MHz.
\end{equation}
The order of this contribution is determined by two small parameters
$\alpha$ and $M_e/M_\mu$ which are written explicitly. The
correction $b_{VP}$ is of the fifth order in $\alpha$ and the first
order in the ratio of the electron and muon masses. The contribution
of the muon vacuum polarization to the hyperfine splitting is
extremely small ($\sim 10^{-6}$ MHz). One should expect that
two-loop vacuum polarization contributions to the hyperfine
structure are suppressed relative to the one-loop VP contribution by
the factor $\alpha/\pi$. This means that at present level of
accuracy we can neglect these corrections because their numerical
value is not exceeding 0.001 MHz. Higher orders of the perturbation
theory which contain one-loop vacuum polarization and the Coulomb
interaction (3) lead to the recoil corrections of order
$\nu_F\alpha\frac{M_e^2}{M_\mu^2}\ln\frac{M_\mu}{M_e}$. Such terms
which can contribute 0.004 MHz are included in the theoretical
error.

Similar contribution to the coefficient $c$ of order $\alpha^6$ can
be found analytically using the potential (19) ($\alpha_1=\alpha
M_e/m_e$):
\begin{equation}
c_{\ VP}=\nu_F\frac{\alpha g_N m_\mu}{6\pi
m_p}\frac{\sqrt{1-\alpha_1^2}(6\alpha_1+\alpha_1^3
-3\pi)+(6-3\alpha_1^2+6\alpha_1^4)\arccos\alpha_1}{3\alpha_1^3
\sqrt{1-\alpha_1^2}}=0.021~MHz.
\end{equation}

Let us consider corrections of the electron vacuum polarization
(15)-(17) in the second order perturbation theory (SOPT) (the
diagram (b) in Fig.1). The contribution of the Coulomb
electron-nucleus interaction (15) to the hyperfine splitting can be
written as follows:
\begin{equation}
b_{\ VP, \ SOPT, \ e-h}=\frac{16\pi\alpha}{3m_em_\mu}\int d{\bf
x}_1\int d{\bf x}_2\int d{\bf x}_3
\frac{\alpha}{3\pi}\int_1^\infty\rho(\xi)d\xi\psi^\ast_{\mu 0}({\bf
x}_3)\psi^\ast_{e 0}({\bf x}_3)\times
\end{equation}
\begin{displaymath}
\times\sum_{n,n'\not =0}^\infty\frac{\psi_{\mu n}({\bf x}_3) \psi_{e
n'}({\bf x}_3)\psi^\ast_{\mu n}({\bf x}_2)\psi^\ast_{e n'}({\bf
x}_1)}{E_{\mu 0}+E_{e0}-E_{\mu n}-E_{en'}}e^{-2m_e\xi x_1}\psi_{\mu
0}({\bf x}_2) \psi_{e0}({\bf x}_1),
\end{displaymath}
where the indices at the coefficient $b$ indicate vacuum
polarization contribution (VP) in the second order PT (SOPT) when
the electron-helion Coulomb VP potential is considered. The
summation in (25) is carried out over the complete system of the
eigenstates of the electron and muon excluding the state with
$n,n'=0$. The computation of the expression (25) is simplified with
the use of the orthogonality condition for the muon wave functions:
\begin{equation}
b_{\ VP, \ SOPT, \ e-h}=\nu_F\frac{32\alpha M_e^2}{3\pi
M_\mu^2}\int_1^\infty \rho(\xi)d\xi\int _0^\infty x_2^3 dx_3\int
_0^\infty x_1 dx_1
e^{-x_1\frac{M_e}{M_\mu}\left(1+\frac{m_e\xi}{\alpha M_e}\right)}
e^{-2x_3\left(1+\frac{M_e}{2M_\mu}\right)}\times
\end{equation}
\begin{displaymath}
\left[\frac{M_\mu}{M_ex_>}-\ln\left(\frac{M_e}{M_\mu}x_<\right)-\ln
\left(\frac{M_e}{M_\mu}x_>\right)+Ei\left(\frac{M_e}{M_\mu}x_<\right)+
\frac{7}{2}-2C-\frac{M_e}{2M_\mu}(x_1+x_3)+
\frac{1-e^{\frac{M_e}{M_\mu}x_<}}{\frac{M_e}{M_\mu}x_<}\right]=
\end{displaymath}
\begin{displaymath}
=0.150~MHz,
\end{displaymath}
where $x_<=\min(x_1,x_3)$, $x_>=\max(x_1,x_3)$, $C=0.577216\ldots$
is the Euler's constant and $Ei(x)$ is the exponential-integral
function. It is necessary to emphasize that the transformation of
the expression (25) into (26) is carried out with the use of the
compact representation for the electron reduced Coulomb Green's
function obtained in Refs.\cite{Hameka}:
\begin{equation}
G_e({\bf x}_1,{\bf x}_3)=\sum_{n\not =0}^\infty\frac{\psi_{en}({\bf
x}_3) \psi_{en}^\ast({\bf x}_1)}{E_{e0}-E_{en}}=-\frac{\alpha
M_e^2}{\pi}e^{-\alpha M_e(x_1+x_3)}\Biggl[\frac{1}{2\alpha M_e x_>}-
\end{equation}
\begin{displaymath}
-\ln(2\alpha M_e x_>)-\ln(2\alpha M_e x_<)+Ei(2\alpha M_e x_<)+
\frac{7}{2}-2C-\alpha M_e(x_1+x_3)+\frac{1-e^{2\alpha M_e
x_<}}{2\alpha M_e x_<}\Biggr].
\end{displaymath}
The contribution (26) has the same order of the magnitude
$O(\alpha^5\frac{M_e}{M_\mu})$ as the previous correction (23) in
the first order perturbation theory. Similar calculation can be
performed in the case of muon-nucleus Coulomb vacuum polarization
potential (16). The intermediate electron state is the 1S state and
the reduced Coulomb Green's function of the system appearing in the
second order PT transforms to the Green's function of the muon. The
correction of the operator (16) to the hyperfine splitting
(coefficient b) is obtained in the following integral form:
\begin{equation}
b_{\ VP, \ SOPT, \ \mu-h}
=\nu_F\frac{\alpha}{3\pi}\int_1^\infty\rho(\xi)d\xi\int_0^\infty
x_3^2dx_3\int_0^\infty x_2dx_2
e^{-x_3\left(1+\frac{M_e}{2M_\mu}\right)}e^{-x_2\left(1+\frac{m_e\xi}
{2M_\mu\alpha}\right)}\times
\end{equation}
\begin{displaymath}
\times\left[\frac{1}{x_>}-\ln x_>-\ln x_<+Ei (x_<)+\frac{7}{2}-2C
-\frac{x_2+x_3}{2}+\frac{1-e^{x_<}}{x_<}\right]= 0.048~MHz.
\end{displaymath}
The vacuum polarization correction to HFS which is determined by the
operator (17) in the second order perturbation theory is the most
difficult for the calculation. Indeed, in this case we have to
consider the intermediate excited states both for the muon and
electron. Following Ref.\cite{LM1} we have divided total
contribution into two parts. The first part in which the
intermediate muon is in the 1S state can be written as:
\begin{equation}
b_{\ VP, \ SOPT, \ \mu-e}(n=0)=\frac{256\alpha^2(\alpha
M_e)^3(2\alpha M_\mu)^3}{9} \int_0^\infty x_3^2dx_3\times
\end{equation}
\begin{displaymath}
\times\int_0^\infty x_1^2 dx_1 e^{-\alpha(M_e+4M_\mu)x_3}\int_1^\infty\rho(\xi)
d\xi\Delta V_{VP~\mu}(x_1)G_e(x_1,x_3),
\end{displaymath}
where the function $V_{VP~\mu}(x_1)$ is equal
\begin{equation}
\Delta V_{VP~\mu}(x_1)=\int d{\bf x}_2 e^{-4\alpha M_\mu
x_2}\frac{(2\alpha M_\mu)^3} {\pi}\frac{\alpha}{|{\bf x}_1-{\bf
x}_2|}e^{-2m_e\xi|{\bf x}_1-{\bf x}_2|}=
\end{equation}
\begin{displaymath}
=\frac{32\alpha^4 M^3_\mu}{x_1(16\alpha^2 M_\mu^2-4m_e^2\xi^2)^2}
\left[8\alpha M_\mu\left(e^{-2m_e\xi x_1}-e^{-4\alpha M_\mu
x_1}\right)+x_1(4m_e^2\xi^2-16\alpha^2 M_\mu^2)e^{-4\alpha M_\mu
x_1}\right].
\end{displaymath}
After the substitution (30) in (29) the numerical integration gives
the result:
\begin{equation}
b_{\ VP, \ SOPT}(n=0)=-0.029~MHz.
\end{equation}
Second part of the vacuum polarization correction to the hyperfine
splitting due to the electron-muon interaction (17) can be presented
as follows:
\begin{equation}
b_{\ VP, \ SOPT, \ \mu e}(n\not
=0)=-\frac{16\alpha^2}{9m_em_\mu}\int d{\bf x}_3\int d{\bf
x}_2\int_1^\infty\rho(\xi)d\xi \psi^\ast_{\mu 0}({\bf
x}_3)\psi^\ast_{e 0}({\bf x}_3)\times
\end{equation}
\begin{displaymath}
\times\sum_{n\not =0}\psi_{\mu n}({\bf x}_3)\psi^\ast_{\mu n}({\bf
x}_2) \frac{M_e}{2\pi}\frac{e^{-{\cal B}|{\bf x}_3-{\bf
x}_1|}}{|{\bf x}_3-{\bf x}_1|}\frac{\alpha}{|{\bf x}_2-{\bf
x}_1|}e^{-2m_e\xi|{\bf x}_2-{\bf x}_1|} \psi_{\mu 0}({\bf
x}_2)\psi_{e 0}({\bf x}_1).
\end{displaymath}
In the expression (32) we have replaced the exact electron Coulomb
Green's function by the free electron Green's function which
contains ${\cal B}=[2M_e(E_{\mu n}-E_{\mu 0}-E_{e 0}]^{1/2}$. (see
more detailed discussion of this approximation in
Refs.\cite{LM1,LM2}). We also replace the electron wave functions by
their values at the origin as in Ref.\cite{LM1} neglecting higher
order recoil corrections. After that the integration over ${\bf
x}_1$ can be done analytically:
\begin{equation}
J=\int d{\bf x}_1\frac{e^{-{\cal B}|{\bf x}_3-{\bf x}_1|}}{|{\bf
x}_3-{\bf x}_1|} \frac{e^{-2m_e\xi|{\bf x}_2-{\bf x}_1|}}{|{\bf
x}_2-{\bf x}_1|}= -\frac{4\pi}{|{\bf x}_3-{\bf x}_2|}\frac{1}{{\cal
B}^2-4m_e^2\xi^2}\left[e^{-{\cal B}|{\bf x}_3-{\bf
x}_2|}-e^{-2m_e\xi|{\bf x}_3-{\bf x}_2|}\right]=
\end{equation}
\begin{displaymath}
=2\pi\Biggl[\frac{\left(1-e^{-2m_e\xi|{\bf x}_3-{\bf x}_2|}\right)}
{2m_e^2\xi^2|{\bf x}_3-{\bf x}_2|}- \frac{{\cal
B}}{2m_e^2\xi^2}+\frac{\left(1-e^{-2m_e\xi|{\bf x}_3-{\bf
x}_2|}\right){\cal B}^2} {8m_e^4\xi^4|{\bf x}_3-{\bf
x}_2|}+\frac{{\cal B}^2|{\bf x}_3-{\bf x}_2|}{4m_e^2\xi^2}-
\end{displaymath}
\begin{displaymath}
-\frac{{\cal B}^3}{8m_e^4\xi^4}-\frac{{\cal B}^3({\bf x}_3-{\bf
x}_1)^2}{12m_e^2\xi^2}+...\Biggr],
\end{displaymath}
where we have performed the expansion of the first exponential in
the square brackets over powers of ${\cal B}|{\bf x}_3-{\bf x}_2|$.
As discussed in Ref.\cite{LM1} one can treat this series as an
expansion over the recoil parameter $\sqrt{M_e/M_\mu}$. For the
further transformation the completeness condition is useful:
\begin{equation}
\sum_{n\not =0}\psi_{\mu n}({\bf x}_3)\psi_{\mu n}^\ast({\bf x}_2)=
\delta({\bf x}_3-{\bf x}_2)-\psi_{\mu 0}({\bf x}_3)\psi_{\mu
0}^\ast({\bf x}_2).
\end{equation}
The wave function orthogonality leads to the zero results for the
second and fifth terms in the square brackets of Eq.(33). The first
term in Eq.(33) gives the leading order contribution in two small
parameters $\alpha$ and $M_e/M_\mu$:
\begin{equation}
b_{\ VP, \ SOPT, \ \mu-e}(n\not =0)=b_{11}+b_{12},~b_{11}=
-\frac{3\alpha^2M_e}{8m_e}\nu_F,
\end{equation}
\begin{equation}
b_{12}=\nu_F\frac{2\alpha^2}{3\pi\frac{m_e}{M_e}}\int_1^\infty\rho(\xi)
\frac{d\xi}{\xi}\frac{M_\mu^4\alpha^4}{(4\alpha
M_\mu+2m_e\xi)^4}\left[256+232\frac{m_e\xi}{M_\mu \alpha}+80\frac{
m_e^2\xi^2}{M_\mu^2\alpha^2}+
10\frac{m_e^3\xi^3}{M_\mu^3\alpha^3}\right].
\end{equation}
The numerical value of the sum $b_{11}+b_{12}=-0.062$ MHz is
included in Table I. It is important to calculate also the
contributions of other terms of the expression (33) to the hyperfine
splitting. Taking the fourth term in Eq.(33) which is proportional
to ${\cal B}^2=2M_e(E_{\mu n}-E_{\mu 0})$ we have performed the
sequence of the transformations in the coordinate representation:
\begin{equation}
\sum_{n=0}^\infty E_{\mu n}\int d{\bf x}_2\int d{\bf x}_3\psi_{\mu
0}^\ast({\bf x}_2) \psi_{\mu n}({\bf x}_3)\psi_{\mu n}^\ast({\bf
x}_2)|{\bf x}_3-{\bf x}_2|\psi_{\mu 0}({\bf x}_2)=
\end{equation}
\begin{displaymath}
=\int d{\bf x}_2\int d{\bf x}_3\delta({\bf x}_3-{\bf x}_2)
\left[-\frac{\nabla^2_3}{2M_\mu}|{\bf x}_3-{\bf x}_2|\psi_{\mu 0}^\ast({\bf x}_3)\right]
\psi_{\mu 0}({\bf x}_2).
\end{displaymath}
Evidently, we have the divergent expression in Eq.(37) due to the
presence of the $\delta$-function. The same divergence occurs in the
other term containing ${\cal B}^2$ in the square brackets of
Eq.(33). But their sum is finite and can be calculated analytically
with the result:
\begin{equation}
b_{{\cal B}^2}=\nu_F\frac{\alpha^2
M_e^2}{m_eM_\mu}\left(18-5\frac{\alpha^2 M_\mu^2}{m_e^2}\right).
\end{equation}
Numerical value of this correction 0.009 MHz is smaller than the
leading order term. Let us consider also the nonzero term in Eq.(33)
proportional to ${\cal B}^3$. First of all, it can be transformed to
the following expression after the integration over $\xi$:
\begin{equation}
b_{{\cal B}^3}=-\nu_F\frac{4\alpha^3}{45\pi}\sqrt{\frac{M_e}{M_\mu}}
\frac{M_e^2}{m_e^2}S_{3/2},
\end{equation}
where the sum $S_{3/2}$ is defined as follows:
\begin{equation}
S_p=\sum_n\left[\left(\frac{E_{\mu n}-E_{\mu
0}}{R_\mu}\right)^{p}\right] |<\psi_{\mu 0}|\frac{{\bf
x}}{a_\mu}|\psi_{\mu n}>|^2,
\end{equation}
$R_\mu=2\alpha^2 M_\mu$, $a_\mu=1/2\alpha M_\mu$. Using the known
analytical expressions for the dipole matrix elements entering in
Eq.(40) in the case of the discrete and continuous spectrum
\cite{HBES,VAF} we can write separately their contributions to the
sum $S_{3/2}$ in the form:
\begin{equation}
S_{3/2}^d=\sum_{n=0}^\infty\frac{2^8n^4(n-1)^{2n-\frac{7}{2}}}{(n+1)^{2n+
\frac{7}{2}}}=1.50989\ldots,
\end{equation}
\begin{equation}
S_{3/2}^c=\int_0^\infty k dk \frac{2^8}{\left(1-e^{-\frac{2\pi}{k}}\right)}
\frac{1}{(1+k^2)^{\frac{7}{2}}}
|\left(\frac{1+ik}{1-ik}\right)^{\frac{i}{k}}|^2=1.76236\ldots  .
\end{equation}
As a result $S_{3/2}=3.2722\ldots$ . The similar calculation of the
sum $S_{1/2}$ relating to this problem (see Ref.\cite{LM1}) gives
$S_{1/2}=2.9380\ldots$. Numerical value (39) is taken into account
in the total result presented in Table I.

\begin{figure}[htbp]
\centering
\includegraphics{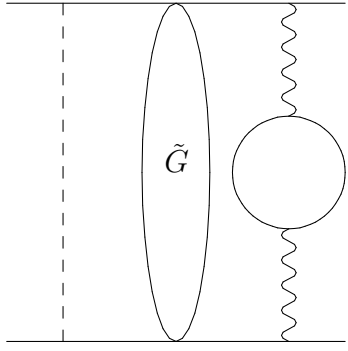}
\caption{Vacuum polarization effects in the second order
perturbation theory. The dashed line represents the first part of
the potential $\Delta H$ (3). The wave line represents the hyperfine
part of the Breit potential.}
\end{figure}

The Coulomb vacuum polarization (15) does not contain the muon
coordinate, so, its contribution to the coefficient $c$ in the
second order PT can be derived taking $n=0$ for the muon state in
the Coulomb Green's function. Moreover, the $\delta({\bf x}_e)$
function in (5) results in the appearance of the electron Green's
function with one zero argument:
\begin{equation}
G_e({\bf x})=\sum_{n\not =0}\frac{\psi_{e~n}(0)\psi^+_{e~n}({\bf
x})}{E_{e~0}-E_{e~n}}=\frac{M_ee^{-\alpha M_e x}}{4\pi x}
\left[4\alpha M_e x(\ln(2\alpha M_e x)+C)+4\alpha^2 M_e^2
x^2-10\alpha M_e x-2\right].
\end{equation}
Corresponding value of the hyperfine splitting is equal
\begin{equation}
c_{\ VP, \ SOPT, \ e-h}=\nu_F\frac{\alpha m_\mu g_N}{2\pi
m_p}\int_1^\infty\rho(\xi)d\xi\frac{2-(1+\frac{\xi}{\alpha_1})
(3+2(1+\frac{\xi}{\alpha_1}))-2(1+\frac{\xi}{\alpha_1})\ln
(1+\frac{\xi}{\alpha_1})}{(1+\frac{\xi}{\alpha_1})^3}=
\end{equation}
\begin{displaymath}
=-0.044~MHz.
\end{displaymath}
The vacuum polarization in the Coulomb $(\mu-N)$ interaction does
not contribute to $c$ in SOPT because of the orthogonality of the
muon wave functions. Let consider correction to the coefficient $c$
arose from (17) in SOPT. Only intermediate muon state with $n=0$ in
the Green's function gives the contribution in this case. Using (43)
we make integration over electron coordinates and present this
correction in the form ($\gamma=m_e\xi/2\alpha M_\mu$,
$\gamma_1=M_e/4M_\mu$):
\begin{equation}
c_{\ VP, \ SOPT,\ e-\mu}=-\nu_F\frac{\alpha m_\mu g_NM_e}{24\pi
m_pM_\mu}\int_1^\infty\rho(\xi)d\xi\frac{2(1-\gamma)^2}{(1-\gamma^2)^2(1+
2\gamma_1)^4(\gamma+2\gamma_1)^3}\times
\end{equation}
\begin{displaymath}
[-\gamma^3(1+2\gamma_1(7+6 \gamma_1))-2\gamma^2 (1 + 3 \gamma_1) (1
+ 2\gamma_1 (7 + 6 \gamma_1))-4\gamma_1^2 [5 + 2\gamma_1 (17 + 12
\gamma_1 (2 + \gamma_1))]-2\gamma \gamma_1\times
\end{displaymath}
\begin{displaymath}
[9 + 2\gamma_1 (37 + 4\gamma_1 (17 + 9 \gamma_1))]+4\gamma_1
(1+2\gamma_1) (\gamma+2\gamma_1)(1+\gamma^2+2 \gamma (1 +
2\gamma_1)+2\gamma_1 (3 + 2\gamma_1))\ln(2\gamma_1)]=
\end{displaymath}
\begin{displaymath}
=0.009~MHz.
\end{displaymath}

There exist another contributions of the second order perturbation
theory in which we have the vacuum polarization perturbation
connected with the hyperfine splitting parts of the Breit potential
(18), (19) (see Fig.2). Other perturbation potential in this case is
determined by the first term of relation (3). We can divide the HFS
correction of (18) into two parts. One part with $n=0$ corresponds
to the ground state muon. The other part with $n\not =0$ accounts
the excited muon states. The $\delta$-function term in Eq.(18) gives
the following contribution to HFS at $n=0$ (compare with
Ref.\cite{LM1}):
\begin{equation}
b_{\ VP, \ SOPT, \
11}(n=0)=\nu_F\frac{\alpha}{3\pi}\int_1^\infty\rho(\xi)d\xi
\frac{11M_e}{16 M_\mu}.
\end{equation}
Obviously, this integral in the variable $\xi$ is divergent. So, we
have to consider the contribution of the second term of the
potential (18) to the hyperfine splitting which is determined by the
more complicated expression:
\begin{equation}
b_{\ VP, \ SOPT, \ 12}(n=0)=\frac{16\alpha^2m_e^2}{9\pi
m_em_\mu}\int_1^\infty \rho(\xi)\xi^2d\xi\int d{\bf x}_3\psi_{e
0}({\bf x}_3)\Delta V_1({\bf x}_3)\times
\end{equation}
\begin{displaymath}
\times\sum_{n'\not =0}
\frac{\psi_{e n'}({\bf x}_3)\psi^\ast_{e n'}({\bf x}_1)}{E_{e0}-E_{en'}}\Delta V_2({\bf x}_1)
\psi_{e 0}({\bf x}_1),
\end{displaymath}
where
\begin{equation}
\Delta V_1({\bf x}_3)=\int d{\bf x}_4\psi^\ast_{\mu 0}({\bf x}_4)
\frac{e^{-2m_e\xi|{\bf x}_3-{\bf x}_4|}}
{|{\bf x}_3-{\bf x}_4|}\psi_{\mu 0}({\bf x}_4)=
\end{equation}
\begin{displaymath}
=\frac{4(2\alpha M_\mu)^3}{x_3[(4\alpha M_\mu)^2-(2m_e\xi)^2]^2}\left[8\alpha M_\mu e^{-2m_e\xi x_3}+
e^{-4\alpha M_\mu x_3}\left(-8\alpha M_\mu-16\alpha^2 M_\mu^2 x_3+4m_e^2\xi^2 x_3\right)\right],
\end{displaymath}
\begin{equation}
\Delta V_2({\bf x}_1)=\int d{\bf x}_2\psi_{\mu 0}({\bf
x}_2)\left(\frac{\alpha}{|{\bf x}_2-{\bf x}_1|}-
\frac{\alpha}{x_1}\right)\psi_{\mu 0}({\bf x}_2)
=-\frac{\alpha}{x_1}(1+2\alpha M_\mu x_1)e^{-4\alpha M_\mu x_1}.
\end{equation}
Nevertheless, integrating in Eq.(47) over all coordinates we obtain
the following result in the leading order in the ratio
$(M_e/M_\mu)$:
\begin{equation}
b_{\ VP, \ SOPT, \ 12}(n=0)=-\nu_F\frac{m_e}{M_e}\frac{M_e^2}{96\pi
M_\mu^2}\int_1^\infty\rho(\xi)\xi
d\xi\frac{32+63\gamma+44\gamma^2+11\gamma^3}{(1+\gamma)^4}.
\end{equation}
This integral also has the divergence at large values of the
parameter $\xi$. But the sum of integrals (46) and (50) is finite:
\begin{equation}
b_{\ VP, \ SOPT, \ 11}(n=0)+b_{\ VP, \ SOPT, \
12}(n=0)=\nu_F\frac{\alpha M_e}{48\pi
M_\mu}\int_1^\infty\rho(\xi)d\xi\frac{11+12\gamma+3\gamma^2}{(1+\gamma)^4}=0.008~MHz.
\end{equation}

Let us consider now the terms in the coefficient $b$ with $n\not
=0$. The delta-like term of the potential (18) gives the
contribution to the HFS known from the calculation of
Ref.\cite{LM1}:
\begin{equation}
b_{\ VP, \ SOPT, \ 21}(n\not =0)=\nu_F\frac{\alpha}{3\pi
M_\mu}\int_1^\infty\rho(\xi)d\xi\left(-\frac{35M_e}{16M_\mu}\right).
\end{equation}
Another correction from the second term of the expression (18) can
be simplified after the replacement the exact electron Green's
function by the free electron Green's function:
\begin{equation}
b_{\ VP, \ SOPT, \ 22}(n\not
=0)=-\frac{16\alpha^3M_em_e^2}{9m_em_\mu}
\int_1^\infty\rho(\xi)\xi^2d\xi\int d{\bf x}_2\int d{\bf x}_3\times
\end{equation}
\begin{displaymath}
\times\int d{\bf x}_4\psi_{\mu 0}^\ast({\bf x}_4)
\frac{e^{-2m_e\xi|{\bf x}_3-{\bf x}_4|}}{|{\bf x}_3-{\bf x}_4|}\sum_{n\not=0}^\infty
\psi_{\mu n}({\bf x}_4)\psi_{\mu n}({\bf x}_2)|{\bf x}_3-{\bf x}_2|\psi_{\mu 0}({\bf x}_2)
\end{displaymath}
The analytical integration in Eq.(53) over all coordinates leads to
the result:
\begin{equation}
b_{\ VP, \ SOPT, \ 22}(n\not =0)=-\nu_F\frac{\alpha M_e}{3\pi
M_\mu}\int_1^\infty\rho(\xi)d\xi\left[
\frac{1}{\gamma}-\frac{1}{(1+\gamma)^4}\left(4+\frac{1}{\gamma}+10\gamma+\frac{215\gamma^2}{16}+
\frac{35\gamma^4}{16}\right)\right].
\end{equation}
The sum of expressions (52) and (54) gives again the finite
contribution to the hyperfine splitting:
\begin{equation}
b_{\ VP, \ SOPT, \ 21}(n\not =0)+\Delta \nu^{hfs}_{VP~SOPT~22}(n\not
=0)=
\end{equation}
\begin{displaymath}
=-\nu_F\frac{\alpha M_e}{3\pi M_\mu}
\int_1^\infty\rho(\xi)d\xi\frac{35+76\gamma+59\gamma^2+16\gamma^3}{16(1+\gamma)^4}=-0.062~MHz.
\end{displaymath}
Despite the fact that the absolute values of the calculated VP
corrections (31), (35), (36), (37), (39), (51), (55) are
sufficiently large , their summary contribution to the hyperfine
splitting (see Table I) is small because they have different signs.

The hyperfine splitting interaction (19) gives the contributions to
the coefficient $c$ in second order PT. Since the muon coordinate
does not enter into the expression (19), we should set $n=0$ for the
muon intermediate states in the Green's function. The initial
formula for this correction is
\begin{equation}
c_{\ VP, \ SOPT}=\frac{8\alpha^3 g_N}{9\pi m_e
m_p}\int_1^\infty\rho(\xi)d\xi\int d{\bf x}_1\int d {\bf x}_3\int
d{\bf x}_4|\psi_{\mu 0}({\bf x}_3)|^2\psi^\ast_{e0}({\bf
x}_4)\psi_{e0}({\bf x}_1)\times
\end{equation}
\begin{displaymath}
\times\left[\frac{1}{|{\bf x}_3-{\bf
x}_4|}-\frac{1}{x_4}\right]G_e({\bf x}_4,{\bf
x}_1)\left(\pi\delta({\bf x}_1)-\frac{m_e^2\xi^2}{x_1}e^{-2m_e\xi
x_1}\right).
\end{displaymath}
The integration over ${\bf x}_3$ can be done analytically as in
(49). Then it is useful to divide (56) into two parts. The
coordinate integration in the first term with the $\delta$ -
function is performed by means of (43). In the second term of (56)
we use the electron Green's function in the form (27). The summary
result can be presented in the integral form in the leading order in
$M_e/M_\mu$:
\begin{equation}
c_{\ VP,\ SOPT}=\frac{2\alpha^5 g_N M_e^4}{9\pi m_e
m_pM_\mu}\int_1^\infty\rho(\xi)d\xi\frac{3+2\frac{m_e\xi}{2\alpha
M_\mu}}{(1+\frac{m_e\xi}{2\alpha M_\mu})^2}=0.013~MHz.
\end{equation}

\section{Nuclear structure and recoil effects}

Another set of significant corrections to the hyperfine splitting of
muonic helium atom which we study in this work is determined by the
nuclear structure and recoil \cite{SGK,GY,BY,M2000}. The charge and
magnetic moment distributions of the helion are described by two
form factors $G_E(k^2)$ and $G_M(k^2)$ for which we use the known
parameterization \cite{mcc}:
\begin{equation}
G_E(k^2)=e^{-\tilde a^2k^2}-\tilde b^2k^2e^{-\tilde c^2k^2}+\tilde
d\left[e^{-\frac{(k+\tilde q_0)^2}{\tilde p^2}}+ e^{-\frac{(k-\tilde
q_0)^2}{\tilde p^2}}\right],
\end{equation}
\begin{equation}
G_M(k^2)=\left[e^{-\tilde a_1^2k^2}-\tilde b_1^2k^2e^{-\tilde
c_1^2k^2}\right],
\end{equation}
where numerous parameters $\tilde a$, $\tilde b$, $\tilde c$,
$\tilde a_1$, $\tilde b_1$, $\tilde c_1$, $\tilde q_0$, $\tilde p$
are written explicitly in \cite{mcc}. In $1\gamma$ - interaction the
nuclear structure correction to the coefficient $c$ is determined by
the amplitudes shown in Fig.3. Purely point contribution in Fig.3(b)
leads to the HFS value (13). Then the nuclear structure correction
is given by
\begin{equation}
c_{\ str,\ 1\gamma}=\nu_F\frac{(1+\kappa_e)g_N
m_\mu}{2m_p}\left[\int G_M(x)e^{-2\alpha M_e x} d{\bf
x}-1\right]=-0.072~MHz.
\end{equation}

\begin{figure}[htbp]
\centering
\includegraphics{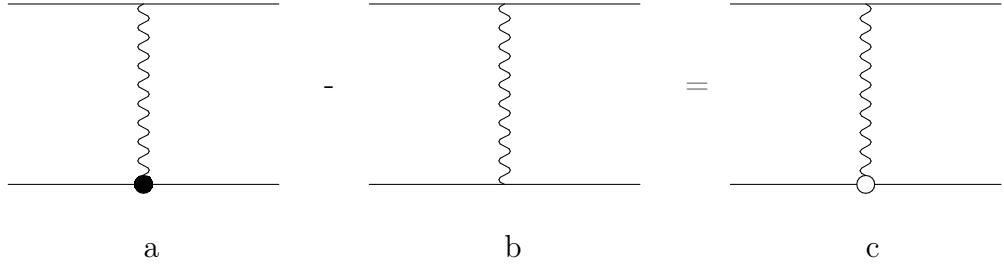}
\caption{Nuclear structure correction to coefficient $c$ in
$1\gamma$ interaction. The bold point represents the nuclear vertex
operator. The wave line represents the hyperfine part of the Breit
potential.}
\end{figure}

\begin{figure}[htbp]
\centering
\includegraphics{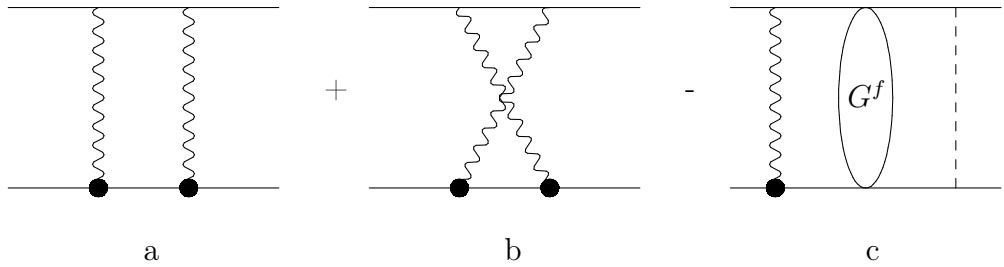}
\caption{Nuclear structure corrections to coefficient $c$ in
$2\gamma$ interactions. The bold point represents the nuclear vertex
operator. The wave line represents the hyperfine part of the Breit
potential. Dash line corresponds to the Coulomb potential.}
\end{figure}

\begin{figure}[htbp]
\centering
\includegraphics{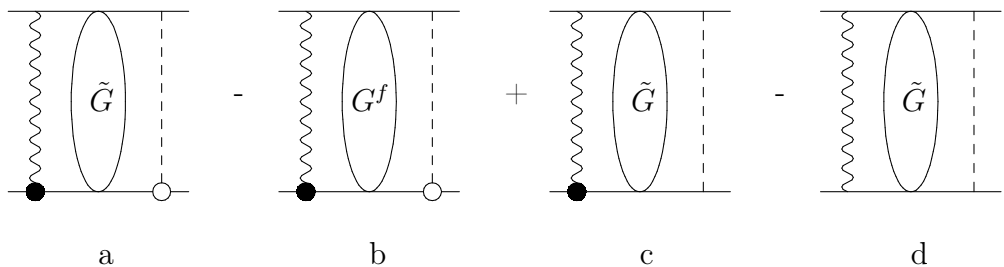}
\caption{Nuclear structure corrections to coefficient $c$ in second
order PT. The bold point represents the nuclear vertex operator. The
wave line represents the hyperfine part of the Breit potential. Dash
line corresponds to the Coulomb potential. $\tilde G$ is the reduced
Coulomb Green's function.}
\end{figure}

Two-photon $(e-h)$ interaction amplitudes (see Fig.4) give the
contribution to HFS of order $\alpha^5$. It can be presented in
terms of the Dirac and Pauli form factors $F_1$ and $F_2$
\cite{M4,M2}:
\begin{equation}
c_{\ str,\ 2\gamma}=\frac{Z^2\alpha^5M_e^3}{3\pi
m_em_h}\delta_{l0}\int_0^\infty\frac{dk}{k}V(k),
\end{equation}
\begin{displaymath}
V(k)=\frac{2F_2^2k^2}{m_em_h}+\frac{M_e}{(m_e-m_h)k(k+\sqrt{4m_e^2+k^2})}
\Biggl[-128F_1^2m_e^2-128F_1F_2m_e^2+16F_1^2k^2+
\end{displaymath}
\begin{displaymath}
+64F_1F_2k^2+16F_2^2k^2+\frac{32F_2^2m_e^2k^2}{m_h^2}+\frac{4F_2^2k^4}
{m_e^2}-\frac{4F_2^2k^4}{m_h^2}\Biggr]+\frac{M_e}{(m_e-m_h)k(k+
\sqrt{4m_h^2+k^2})}\times
\end{displaymath}
\begin{displaymath}
\times\left[128F_1^2m_h^2+128F_1F_2m_h^2-16F_1^2k^2-64F_1F_2k^2-48F_2^2k^2\right].
\end{displaymath}
The subtraction term in Fig.4 is taken as follows:
\begin{equation}
c_{\ iter,\ str}=\frac{64}{3}\frac{M_e^4Z^2\alpha^5
g_N}{6m_em_h\pi}\int_0^\infty\frac{dk}{k^2}G_M(k^2).
\end{equation}
It is only the part of the iteration contribution
$<V_{1\gamma}\times G^f\times V_{1\gamma}>_{str}^{hfs}$. Other parts
are used in the second order PT (see Fig.5). As a result we obtain:
\begin{equation}
c_{\ str,\ 2\gamma}+c_{\ iter,\ str}=-0.077~MHz.
\end{equation}

\begin{figure}[htbp]
\centering
\includegraphics{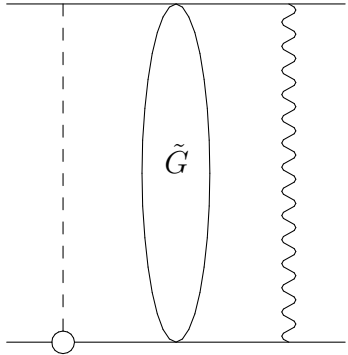}
\caption{Nuclear structure correction to coefficient $b$ in the
second order perturbation theory. The wave line represents the
hyperfine $(e-\mu)$ interaction. $\tilde G$ is the reduced Coulomb
Green's function.}
\end{figure}

The nuclear structure corrections to the coefficient $c$ in second
order PT which are presented in Fig.5. Here we have two different
contributions. First contribution is related with diagrams in
Fig.5(a),(b) when the hyperfine part of one perturbation is
determined by magnetic form factor $G_M$ and the other perturbation
is described by the charge radius of the helion $r_h$:
\begin{equation}
\Delta V^C_{str, e-h}({\bf r})=\frac{2}{3}\pi
Z\alpha<r^2_h>\delta({\bf r}).
\end{equation}
The general integral structure of this correction and its numerical
value are the following:
\begin{equation}
c_{1, \ str, \ SOPT, e-h}=\nu_F\frac{4\pi Zg_N m_\mu}{6\alpha M_e
m_p}<r^2_h>\int_0^\infty x dx e^{-2x}[4x(\ln
2x+C)+4x^2-10x]G_M(\frac{x}{\alpha M_e})=
\end{equation}
\begin{displaymath}
=0.00005~MHz.
\end{displaymath}
Numerical value of the contribution $c_{1, \ str, \ SOPT, e-h}$ is
obtained by means of the charge radius of the helion
$r_h=1.9642(11)$ fm \cite{Drake}. The second nuclear structure
contribution from the diagrams Fig.5 (c), (d) is evaluated by means
of the potential $\Delta H$ (3) and the helion magnetic form factor.
For the amplitude in Fig.5(c) we make the integration over the muon
coordinate in the muon state at $n=0$ and present the correction to
the coefficient $c$ in the form:
\begin{equation}
c_{2, \ str, \ SOPT, e-h}+c_1=\frac{\pi \alpha
M_e^5g_N(1+\kappa_e)}{24 m_em_pM_\mu^5}\int_0^\infty x^2
e^{-\frac{M_e}{2M_\mu}x}G_M(\frac{x}{4\alpha M_\mu})dx\int_0^\infty
y (1+\frac{y}{2})e^{-y(1+\frac{M_e}{2M_\mu})}dy \times
\end{equation}
\begin{displaymath}
\left[\frac{1}{\frac{M_e}{2M_\mu}x_>}-\ln(\frac{M_e}{2M_\mu}x_>)-
\ln(\frac{M_e}{2M_\mu}x_<)+Ei(\frac{M_e}{2M_\mu}x_<)+\frac{7}{2}-2C-\frac{M_e}{2M_\mu}
(x+y)+\frac{1-e^{\frac{M_e}{2M_\mu}x_<}}{\frac{M_e}{2M_\mu}x_<}\right]=
\end{displaymath}
\begin{displaymath}
=8.249~MHz.
\end{displaymath}
Subtracting the point contribution $c_1$ (14) we find $c_{2, \ str,
\ SOPT, e-h}=-0.074$ MHz.

There is the nuclear structure contribution to the coefficient $b$
in second order PT which is presented in Fig.6. If we consider the
Coulomb interaction between the muon and helion, then the structure
correction takes on form:
\begin{equation}
b_{str,\mu-h}=\frac{64\pi^2\alpha^2}{9m_em_\mu}r_h^2
\frac{1}{\sqrt{\pi}}\left(2\alpha M_\mu\right)^{3/2} \int d{\bf
x}_3\psi^\ast_{\mu 0}({\bf x}_3)|\psi_{e0}({\bf x}_3)|^2G_\mu({\bf
x}_3,0,E_{\mu 0}).
\end{equation}
After that the analytical integration over the coordinate ${\bf
x}_3$ in Eq.(67) can be carried out using the representation of the
muon Green's function similar to expression (43). The result of the
integration of order $O(\alpha^6)$ is written as an expansion in the
ratio $M_e/M_\mu$:
\begin{equation}
b_{str,\mu-h}=-\nu_F\frac{8}{3}\alpha^2M_\mu^2
r_\alpha^2\left(3\frac{M_e}{M_\mu}-
\frac{11}{2}\frac{M_e^2}{M_\mu^2}+\ldots\right)=-0.010 ~MHz.
\end{equation}
The same approach can be used in the calculation of the
electron-nucleus interaction. The electron feels as well the
distribution of the helion electric charge. The corresponding
contribution of the nuclear structure effect to the hyperfine
splitting is determined by the expression:
\begin{equation}
b_{str,e-h}=\frac{64\pi^2\alpha^2}{9m_em_\mu}r_h^2 \int d{\bf x}_1
\int d{\bf x}_3|\psi^\ast_{\mu 0}({\bf x}_3)|^2\psi_{e0}({\bf
x}_3)G_\mu({\bf x}_3,{\bf x}_1,E_{e 0}) \psi_{e0}({\bf
x}_1)\delta({\bf x}_1).
\end{equation}
Performing the analytical integration in Eq.(69) we obtain the
following series:
\begin{equation}
b_{str,e-h}=-\nu_F\frac{4}{3}\alpha^2
M_e^2r_\alpha^2\left[5-\ln\frac{M_e}{M_\mu}+
\frac{M_e^2}{M_\mu^2}\left(3\ln\frac{M_e}{M_\mu}-7\right)+\frac{M_e^2}{M_\mu^2}\left(\frac{17}{2}-
3\ln\frac{M_e}{M_\mu}\right)\ldots\right]=
\end{equation}
\begin{displaymath}
=-0.003~MHz.
\end{displaymath}
We have included in Table I the total nuclear structure contribution
to the coefficient $b$ which is equal to the sum of the numerical
values (68) and (70).

\begin{figure}[htbp]
\centering
\includegraphics{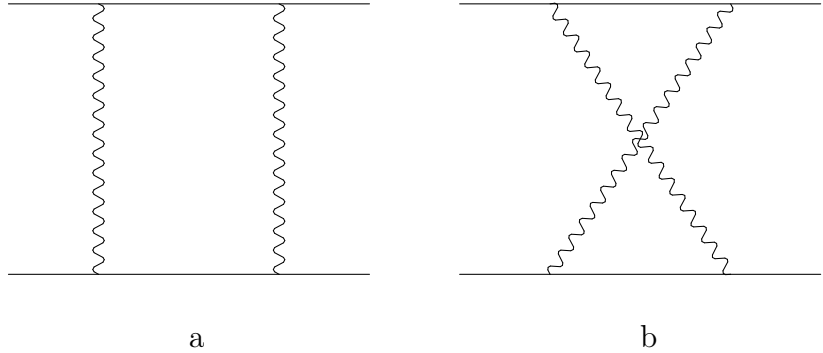}
\caption{Two photon exchange amplitudes in the electron-muon hyperfine
interaction.}
\end{figure}

Special attention has to be given to the recoil corrections
connected with two-photon exchange diagrams shown in Fig.7 in the
case of the electron-muon interaction.  The leading order recoil
contribution to the interaction operator between the muon and
electron is determined by the expression \cite{EGS,Chen1,RA}:
\begin{equation}
\Delta V_{rec,\mu -e}^{hfs}({\bf x}_{\mu
e})=-8\frac{\alpha^2}{m_\mu^2-m_e^2}\ln\frac{m_\mu}{m_e}({\bf s}_\mu
{\bf s}_e) \delta({\bf x}_{\mu e}).
\end{equation}
Averaging the potential $\Delta V_{rec,\mu -e}^{hfs}$ over the wave
functions (4) we obtain the recoil correction to the coefficient
$b$:
\begin{equation}
b_{rec,\mu-e}=\nu_F\frac{3\alpha}{\pi}\frac{m_em_\mu}{m_\mu^2-m_e^2}\ln\frac{m_\mu}
{m_e}=0.812~MHz.
\end{equation}
There exist also the two-photon interactions between the bound
particles of muonic helium atom when one hyperfine photon transfers
the interaction from the electron to muon and another Coulomb photon
from the electron to the nucleus (or from the muon to the nucleus).
Supposing that these amplitudes give smaller contribution to the
hyperfine splitting we included them in the theoretical error.

\section{Electron vertex corrections}

In the initial approximation the potential of the hyperfine
splitting is determined by Eq.(5). It leads to the energy splitting
of order $\alpha^4$. In QED perturbation theory there is the
electron vertex correction to the potential (5) which is defined by
the diagram in Fig.8(a). In momentum representation the
corresponding operator of hyperfine interaction has the form:
\begin{equation}
\Delta V^{hfs}_{vertex}(k^2)=-\frac{8\alpha^2}{3m_em_\mu}
\left(\frac{{\mathstrut\bm\sigma}_e{\mathstrut\bm\sigma}_\mu}{4}\right)
\left[G_M^{(e)}(k^2)-1\right],
\end{equation}
where $G_M^{(e)}(k^2)$ is the electron magnetic form factor. We
extracted for the convenience the factor $\alpha/\pi$ from
$\left[G_M^{(e)}(k^2)-1\right]$. Usually used approximation for the
electron magnetic form factor $G_M^{(e)}(k^2)\approx
G_M^{(e)}(0)=1+\kappa_e$ is not quite correct in this task. Indeed,
characteristic momentum of the exchanged photon is $k\sim\alpha
M_\mu$. It is impossible to neglect it in the magnetic form factor
as compared with the electron mass $m_e$. So, we should use exact
one-loop expression for the magnetic form factor which was obtained
by many authors \cite{t4}. Let us note that the Dirac form factor of
the electron is dependent on the parameter of the infrared cutoff
$\lambda$. We take it in the form $\lambda=m_e\alpha$ using the
prescription $m_e\alpha^2\ll\lambda\ll m_e$ from Ref.\cite{HBES}.

\begin{figure}[htbp]
\centering
\includegraphics{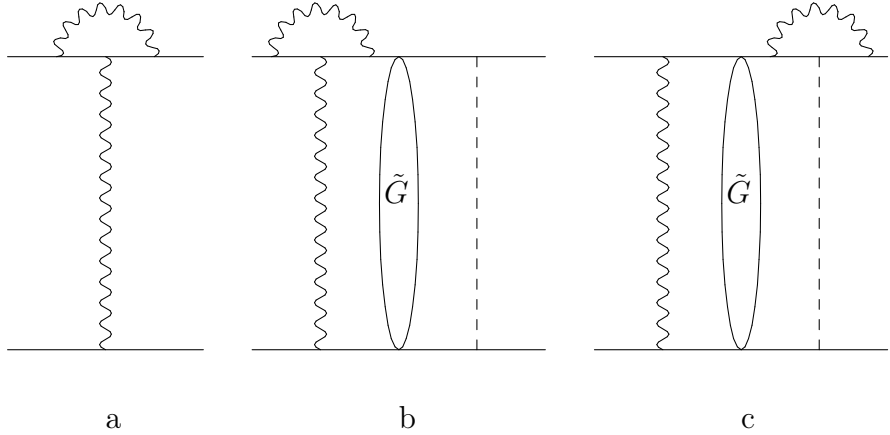}
\caption{The electron vertex corrections. The dashed line represents
the Coulomb photon. The wave line represents the hyperfine part of
the Breit potential. $\tilde G$ is the reduced Coulomb Green's
function.}
\end{figure}

Using the Fourier transform of the potential (73) and averaging the
obtained expression over wave functions (4) we represent the
electron vertex correction to the hyperfine splitting as follows:
\begin{equation}
b_{vertex,\
1\gamma}=\nu_F\frac{\alpha}{32\pi^2}\left(\frac{M_e}{M_\mu}\right)
\left(\frac{m_e}{\alpha M_\mu}\right)^3\int_0^\infty
k^2dk\left[G_M^{(e)}(k^2)-1\right]\times
\end{equation}
\begin{displaymath}
\times\left\{\left[1+\left(\frac{m_e}{4\alpha
M_\mu}\right)^2k^2\right]\left[\left(\frac{M_e}{2M_\mu}\right)^2+\left(\frac{m_e}{4\alpha
M_\mu}\right)^2k^2\right]^2\right\}^{-1}=4.218~MHz.
\end{displaymath}
Let us remark that the contribution (74) is of order $\alpha^5$.
Numerical value (74) is obtained after numerical integration with
the one-loop expression of the electron magnetic form factor
$G_M^{(e)}(k^2)$. If we use the value $G_M^{(e)}(k^2=0)$ then the
electron vertex correction is equal $5.266$ MHz. So, using the exact
expression of the electron form factors in the one-loop
approximation we observe the 1 MHz decrease of the vertex correction
to the hyperfine splitting from $1\gamma$ interaction. Taking the
expression (73) as an additional perturbation potential we have to
calculate its contribution to HFS in the second order perturbation
theory (see the diagram in Fig.8(b)). In this case the dashed line
represents the Coulomb Hamiltonian $\Delta H$ (3). Following the
method of the calculation formulated in previous section (see also
Refs.\cite{LM1,LM2}) we divide again total contribution from the
amplitude in Fig.8(b) into two parts which correspond to the muon
ground state $(n=0)$ and muon excited intermediate states $(n\not
=0)$. In this way the first contribution with $n=0$ takes the form:
\begin{equation}
b_{vertex, \
SOPT}(n=0)=\frac{8\alpha^2}{3\pi^2m_em_\mu}\int_0^\infty
k\left[G_M^{(e)}(k^2)-1\right]dk\int d{\bf x}_1\int d{\bf
x}_3\psi_{e0}({\bf x}_3)\times
\end{equation}
\begin{displaymath}
\times\Delta\tilde V_1(k,{\bf x}_3)G_e({\bf x}_1,{\bf x}_3)\Delta
V_2({\bf x}_1)\psi_{e0}({\bf x}_1),
\end{displaymath}
where $\Delta V_2({\bf x}_1)$ is defined by Eq.(49) and
\begin{equation}
\Delta \tilde V_1(k,{\bf x}_3)=\int d{\bf x}_4\psi_{\mu 0}({\bf
x}_4)\frac{\sin(k|{\bf x}_3-{\bf x}_4|)}{|{\bf x}_3-{\bf
x}_4|}\psi_{\mu 0}({\bf x}_4)=\frac{\sin\left(\frac{kx_3}{4\alpha
M_\mu}\right)}{x_3}\frac{1}{\left[1+\frac{k^2}{(4\alpha
M_\mu)^2}\right]^2}.
\end{equation}
Substituting the electron Green's function (27) in Eq.(75) we
transform desired relation to the integral form:
\begin{equation}
b_{vertex, \ SOPT}(n=0)=\nu_F\frac{\alpha}{16\pi^2}\left(\frac{m_e}
{\alpha M_\mu}\right)^2\left(\frac{M_e}{M_\mu}\right)^2\int_0^\infty
\frac{k\left[G_M^{(e)}(k^2)-1\right]dk}{\left[1+\frac{m_e^2k^2}{(4\alpha
M_\mu)^2}\right]^2}\times
\end{equation}
\begin{displaymath}
\times\int_0^\infty x_3e^{-\frac{M_e}{2M_\mu}x_3}\sin\left(\frac{m_e
k}{4\alpha M_\mu}x_3\right)dx_3\int_0^\infty
x_1\left(1+\frac{x_1}{2}\right)e^{-x_1\left(1+\frac{M_e}{2M_\mu}\right)}dx_1
\times
\end{displaymath}
\begin{displaymath}
\left[\frac{2M_\mu}{M_ex_>}-\ln(\frac{M_e}{2M_\mu}x_<)-\ln
(\frac{M_e}{2M_\mu}x_>)+Ei(\frac{M_e}{2M_\mu}x_<)+
\frac{7}{2}-2C-\frac{M_e}{4M_\mu}(x_1+x_3)+
\frac{1-e^{\frac{M_e}{2M_\mu}x_<}}{\frac{M_e}{2M_\mu}x_<}\right]
\end{displaymath}
\begin{displaymath}
=-0.208~MHz.
\end{displaymath}
One integration over the coordinate $x_1$ is carried out
analytically and two other integrations are performed numerically.
Second part of the vertex contribution (Fig.8(b)) with $n\not =0$
can be reduced to the following form after several simplifications
which are discussed in section II (see also Refs.\cite{LM1,LM2}):
\begin{equation}
b_{vertex,\ SOPT}(n\not
=0)=\nu_F\frac{8\alpha^4M_eM_\mu^3}{\pi^3}\int e^{-2\alpha M_\mu
x_2}d{\bf x}_2\int e^{-\alpha M_e x_3}d{\bf x}_3\int e^{-2\alpha
M_\mu x_4}d{\bf x}_4\times
\end{equation}
\begin{displaymath}
\times\int_0^\infty k\sin(k|{\bf x}_3-{\bf
x}_4|)\left(G_M^{(e)}(k^2)-1\right)\frac{|{\bf x}_3-{\bf
x}_2|}{|{\bf x}_3-{\bf x}_4|}\left[\delta({\bf x}_4-{\bf x}_2)-
\psi_{\mu 0}({\bf x}_4)\psi_{\mu 0}({\bf x}_2)\right].
\end{displaymath}
We divide expression (78) into two parts as provided by two terms in
the square brackets of (78). After that the integration (78) over
the coordinates ${\bf x}_1$, ${\bf x}_3$ is carried out
analytically. In the issue we obtain ($\gamma_2=m_ek/4\alpha
M_\mu$):
\begin{equation}
b_{1, vertex,\
SOPT}(n\not=0)=\nu_F\frac{\alpha}{32\pi^2}\left(\frac{m_e} {\alpha
M_\mu}\right)^3\frac{M_e}{M_\mu}\int_0^\infty
k^2\left[G_M^{(e)}(k^2)-1\right]dk\frac{1}{(\gamma_1^2-1)^3}\times
\end{equation}
\begin{displaymath}
\times\left[\frac{4\gamma_1(\gamma_1^2-1)}{(1+\gamma_2^2)^3}-\frac{\gamma_1
(3+\gamma_1^2)}{(1+\gamma_2^2)^2}+\frac{4\gamma_1^2(\gamma_1^2-1)}{(\gamma_1^2+
\gamma_2^2)^3}+\frac{1+3\gamma_1^2}{(\gamma_1^2+\gamma_2^2)^2}\right]=2.528~MHz,
\end{displaymath}

\begin{equation}
b_{2, vertex,\
SOPT}(n\not=0)=-\nu_F\frac{\alpha}{32\pi^2}\left(\frac{m_e} {\alpha
M_\mu}\right)^3\frac{M_e}{M_\mu}\int_0^\infty
k^2\left[G_M^{(e)}(k^2)-1\right]dk\times
\end{equation}
\begin{displaymath}
\times\frac{1}{(1+\gamma_2^2)^2}\left[\frac{2}{(\gamma_1^2+\gamma_2^2)}-\frac{(\gamma_1+1)}
{[(1+\gamma_1)^2+\gamma_2^2]^2}-\frac{2}{(\gamma_1+1)^2+
\gamma_2^2}-\frac{\gamma_2^2-3\gamma_1^2}{(\gamma_1^2+\gamma_2^2)^3}\right]=-0.831~MHz.
\end{displaymath}
It is necessary to emphasize that the theoretical error in the
contributions $b_{1,2,vertex, SOPT}(n\not =0)$ is determined by the
factor $\sqrt{M_e/M_\mu}$ connected with the omitted terms of the
expansion similar to Eq.(33) (see also Refs.\cite{LM1,LM2}). It can
amount to $10\%$ of the results (79), (80) that is the value near
0.2 MHz.

Until now we consider the electron vertex corrections connected with
the hyperfine part of the interaction Hamiltonian (5). But in the
second order perturbation theory we should analyze vertex
corrections to the Coulomb interactions of the electron and muon,
electron and nucleus. Then in the coordinate representation we have
the following potential:
\begin{equation}
\Delta V^C_{vertex, eN}(x_e)+\Delta
V^C_{vertex,e\mu}(x_{e\mu})=\frac{2\alpha^2}{\pi^2}\int_0^\infty
\frac{\left[G_E^{(e)}(k^2)-1\right]}{k}dk\left(\frac{\sin(kx_{e\mu})}{x_{e\mu}}-
2\frac{\sin(kx_e)}{x_e}\right),
\end{equation}
where we extract again the factor $\alpha/\pi$ from
$\left[G_E^{(e)}(k^2)-1\right]$. $G_E^{(e)}$ is the electron
electric form factor. One part of the contribution in Fig.8(c) is
specified by the electron-muon intermediate states in which the muon
is in the ground state $n=0$. This correction is determined by both
terms in large parentheses of Eq.(81) and can be presented as
follows:
\begin{equation}
b_{C, vertex,\ SOPT}(n=0)=\nu_F\frac{\alpha}{\pi^2}
\left(\frac{M_e}{M_\mu}\right)^2\int_0^\infty x_3^2
e^{-x_3\left(1+\frac{M_e}{2M_\mu}\right)}dx_3\times
\end{equation}
\begin{displaymath}
\times\int_0^\infty x_1e^{-\frac{M_e}{2M_\mu}x_1}dx_1 \int_0^\infty
\frac{\left[G_E^{(e)}(k^2)-1\right]dk}{k}\sin\left(\frac{m_ek}{4\alpha
M_\mu}x_1\right)\left\{1-\frac{1}{2\left[\frac{m_e^2k^2}{(4\alpha
M_\mu)^2}+1 \right]^2}\right\}\times
\end{displaymath}
\begin{displaymath}
\left[\frac{2M_\mu}{M_ex_>}-\ln(\frac{M_e}{2M_\mu}x_<)-\ln
(\frac{M_e}{2M_\mu}x_>)+Ei(\frac{M_e}{2M_\mu}x_<)+
\frac{7}{2}-2C-\frac{M_e}{4M_\mu}(x_1+x_3)+
\frac{1-e^{\frac{M_e}{2M_\mu}x_<}}{\frac{M_e}{2M_\mu}x_<}\right]
\end{displaymath}
\begin{displaymath}
=-1.303~MHz.
\end{displaymath}
The index "C" means that the vertex correction to the Coulomb part
of the Hamiltonian is considered. Excited states of the muon $(n\not
=0)$ contribute to another part of the matrix element (Fig.8(c)).
Changing the Coulomb Green's function of the electron by free
Green's function (see discussion in section II) we can carry out the
coordinate integration and express the correction to HFS as
one-dimensional integral:
\begin{equation}
b_{C, vertex,\ SOPT}(n\not =
0)=-\nu_F\frac{8\alpha}{\pi^2}\frac{M_e}{M_\mu}\left(\frac{\alpha
M_\mu}{m_e}\right)\int_0^\infty
\frac{\left[G_E^{(e)}(k^2)-1\right]dk}{k^2}
\left\{1-\frac{1}{\left[1+\frac{m_e^2k^2}{(4\alpha M_\mu)^2}
\right]^4}\right\}=
\end{equation}
\begin{displaymath}
=1.806~MHz.
\end{displaymath}
The electron vertex corrections investigated in this section have
the order $\alpha^5$ in the hyperfine interval. Summary value of all
obtained contributions (74), (77), (79), (80), (82), (83) is equal
to 6.210 MHz. It differs by a significant value 0.944 MHz from the
result 5.266 MHz which was used previously by many authors for the
estimation of the electron anomalous magnetic moment contribution.
On our opinion, it is necessary to use the same approach for the
calculation of the electron vertex corrections by the variational
method \cite{VK2000} in which the bound state wave function has the
form $\psi({\bf x}_e,{\bf x}_\mu,{\bf
x}_{e\mu})=\sum_iC_ie^{-\alpha_ix_e-\beta_ix_\mu-\gamma_ix_{e\mu}}$
and the parameters $\alpha_i$, $\beta_i$, $\gamma_i$ are chosen
randomly between some minimal and maximal values.

\begin{table}
\caption{\label{t1}Hyperfine splitting of the ground state in the
muonic helium atom $(\mu\ e \ ^3_2He)$.}
\bigskip
\begin{ruledtabular}
\begin{tabular}{|c|c|c|c|}  \hline
Contribution to the HFS & $b$, MHz & $c$, MHz & Reference   \\
\hline The Fermi splitting &4516.307& 1083.256 & (11), (13), \cite{LM3,LM1,LM2}  \\
\hline
Recoil correction of order  & -64.322& 8.323&  (12), (14), \cite{LM3}  \\
$\alpha^4(m_e/m_\mu)$    &       &   &   \\    \hline Correction of
muon anomalous &  5.266 &---  &  (11), \cite{LM3,LM1}  \\
magnetic moment of order $\alpha^5$&  &  &   \\  \hline
Recoil correction of order &   0.105 & --- & \cite{LM1,LM2}   \\
$\alpha^4(M_e/m_\alpha)\sqrt{(M_e/M_\mu)}$   &   &      &  \\
\hline
Relativistic correction of order $\alpha^6$  & 0.040 &0.087   & \cite{HH1,Chen1} \\
\hline
One-loop VP contribution in $1\gamma$&  0.036 & 0.021 &   (23), (24)    \\
interaction of orders $\alpha^5,\alpha^6$  &  &   &  \\
\hline
One-loop VP contribution in the & 0.062  &-0.023  & (26),(28),(31),(35),(36),(38),    \\
second order PT&    &     &  (39),(44),(45),(51),(55),(57)    \\
\hline
Nuclear structure correction & ---   &  -0.072    &  (60)     \\
in $1\gamma$ interaction of order $\alpha^6$    &    &      &       \\
\hline
Nuclear structure and recoil correction& ---   &  -0.077    & (63)      \\
in $2\gamma$ interactions of order $\alpha^5$    &    &      &       \\
\hline
Nuclear structure correction of order& -0.013   &  -0.074    & (14),(65),(66)      \\
$\alpha^6$ in second order PT     &    &      &       \\
\hline
Recoil correction of order & 0.812  &---  & (72) ,\cite{Chen1,RA} \\
$\alpha^5(m_e/m_\mu)\ln(m_e/m_\mu)$    &   &   &    \\   \hline
Electron vertex
correction of order $\alpha^6$ &  -0.615 & -0.035&   \cite{BE,KP,KKS,EGS}  \\
\hline
Electron vertex contribution &  6.210& --- & (74),(77),(79),    \\
of order $\alpha^5$ &       &  & (80),(82),(83)    \\
\hline
Summary contribution &  4463.888 &  1091.406  &  $\Delta\nu^{hfs}=4166.471$~MHz\\
\hline
\end{tabular}
\end{ruledtabular}
\end{table}

\section{Conclusions}

In the present study, we have performed the analytical and numerical
calculation of several important contributions to the hyperfine
splitting of the ground state in muonic helium atom connected with
the vacuum polarization, the nuclear structure, recoil effects and
the electron vertex corrections. To solve this task we use the
method of the perturbation theory which was formulated previously
for the description of the muonic helium hyperfine splitting in
Refs.\cite{LM3,LM1,LM2}. We have considered corrections of the
electron vacuum polarization, electron electromagnetic form factors
and the nuclear structure effects of orders $\alpha^5$ and
$\alpha^6$. The numerical values of the corresponding contributions
are displayed in Table I. We present in Table I the references to
the calculations of other corrections which are not considered here.
The relativistic correction was obtained in Ref.\cite{HH2,Chen1},
the electron vertex corrections of order $\alpha(Z\alpha)E_F$ were
calculated in the case of hydrogenic atoms in
Refs.\cite{EGS,BE,KP,KKS}. Basic contributions to the hyperfine
splitting obtained by Lakdawala and Mohr are also included in Table
I because our calculation is closely related to their approach.

Let us list basic points related to the calculation.

1. For muonic helium atom, the vacuum polarization effects are
important for obtaining the high accuracy of the calculation. They
give rise to the modification of the two-particle interaction
potential which provides the $\alpha^5\frac{M_e}{M_\mu}$-order
corrections to the hyperfine structure. The next to leading order
vacuum polarization corrections (two-loop vacuum polarization) are
negligible.

2. The electron vertex corrections to the coefficient $b$ should be
considered with the exact account of the one-loop electromagnetic
form factors of the electron because the characteristic momentum
incoming in the electron vertex operator is of order of the electron
mass.

3. The nuclear structure corrections to the ground state hyperfine
splitting are expressed in terms of electromagnetic form factors and
the charge radius of the helion.

4. Analyzing the one-loop electron vacuum polarization and vertex
effects and the nuclear structure contributions in each order in
$\alpha$, we have taken into account recoil terms proportional to
the ratio of the electron and muon masses.

The resulting numerical value 4466.471 MHz of the smaller ground
state hyperfine splitting in muonic helium $(\mu\ e \ ^3_2He)$ is
presented in Table I. It is sufficiently close both to the
experimental result (1) and the earlier performed calculations by
the perturbation theory, variational approach in
\cite{LM3,HH1,Chen1}. The estimation of the theoretical uncertainty
can be done in terms of the Fermi energy $\nu_F$  and small
parameters $\alpha$ and the ratio of the particle masses. On our
opinion, there exist several main sources of the theoretical errors.
First of all, as we mentioned above comprehensive analytical and
numerical calculation of recoil corrections of orders
$\alpha^4\frac{M_e}{M_\mu}$, $\alpha^4\frac{M^2_e}{M^2_\mu}$,
$\alpha^4\frac{M^2_e}{M^2_\mu}\ln(M_\mu/M_e)$ was carried out by
Lakdawala and Mohr in the second order PT in Refs.\cite{LM1,LM2}.
The error of their calculation connected with the correction
$\nu_F\frac{M_e^2}{M_\mu^2}\ln\frac{M_\mu}{M_e}$ consists 0.6 MHz.
The second source of the error is related to contributions of order
$\alpha^2 \nu_F\approx 0.2$ MHz which appear both from QED
amplitudes and in higher orders of the perturbation theory. Another
part of the theoretical error is determined by the two-photon
three-body exchange amplitudes mentioned above. They are of the
fifth order over $\alpha$ and contain the recoil parameter
$(m_e/m_\alpha)\ln(m_e/m_\alpha)$, so that their possible numerical
value can be equal $\pm 0.05$ MHz. Finally, a part of theoretical
error is connected with our calculation of the electron vertex
corrections of order $\alpha^5$ in section IV. It consists at least
0.2 MHz (see the discussion after Eq.(80)). We neglect also the
electron vertex contributions of order $\nu_F\alpha M_e/M_\mu\approx
0.2$ MHz which appear in higher orders of the perturbation theory.
Thereby, the total theoretical uncertainty is not exceeded $\pm 0.7$
MHz. The existing difference between the obtained theoretical result
and experimental value of the hyperfine splitting (1) equal to 0.171
MHz lies in the range of total error. Theoretical error which
remains sufficiently large in the comparison with the experimental
uncertainty, initiates further theoretical investigation of the
higher order recoil contributions and more careful construction of
the three-particle interaction operator connected with the
multiphoton exchanges.

\begin{acknowledgments}

This work was supported by the Federal Program "Scientific and
pedagogical personnel of innovative Russia under Grant No. NK-20P/1.
\end{acknowledgments}


\begin{thebibliography}{99}
\bibitem{LM3}S. D. Lakdawala and P. Mohr, Phys. Rev. A {\bf 24},
2224 (1981).
\bibitem{HH1}K.-N. Huang and V. W. Hughes, Phys. Rev. A {\bf 26}, 2330 (1982).
\bibitem{HH2}K.-N. Huang and V. W. Huges, Phys. Rev. A {\bf 20}, 706
(1979).
\bibitem{RD1}R. L. Drachman, J. Phys. B {\bf 16}, L749 (1983).
\bibitem{Chen1}M.-K. Chen, J. Phys. B {\bf 26}, 2263 (1993).
\bibitem{Chen2}M.-K. Chen, J. Phys. B {\bf 23}, 4041 (1990).
\bibitem{Amusia1}V. L. Yakhontov and M. Ya. Amusia, J. Phys. B {\bf
27}, 3743 (1994)
\bibitem{Krivec}R. Krivec and V. B. Mandelzweig, Phys. Rev. {\bf 57}, 4976 (1998).
\bibitem{AF}A. M. Frolov, Phys. Rev. A {\bf 61}, 022509 (2000).
\bibitem{Nature}R. Pohl, A. Antognini, F. Nez et al., Nature {\bf
466}, 213 (2010).
\bibitem{Gladish}M. Gladish et al., Proc. 8th Int. Conf. on Atom.
Phys., ed. I. Lindgren, A. Rosen and S. Svanberg, Plenum, New York,
1983.
\bibitem{HBES}H. A. Bethe and E. E. Salpeter, Quantum mechanics of one-
and two-electron atoms, Springer, Berlin, 1957.
\bibitem{EGS}M. I. Eides, H. Grotch and V. A. Shelyuto, Phys. Rep. {\bf 342},
62 (2001).
\bibitem{SGK}S. G. Karshenboim, Phys. Rep. {\bf 422}, 1 (2005).
\bibitem{M2004}R. N. Faustov and A. P. Martynenko, Yad. Fiz. {\bf 45}, 770 (1987)
[Sov. J. Nucl. Phys. {\bf 45}, 479 (1987)]; Phys. Atom. Nucl. {\bf
66}, 1719, (2003); JETP {\bf 88}, 672, (1999); JETP {\bf 98}, 39
(2004)
\bibitem{M3}A. P. Martynenko, Phys. Rev. A {\bf 76}, 012505 (2007).
\bibitem{M4}A. P. Martynenko, JETP {\bf 106}, 691 (2008); arXiv:0710.3237
[hep-ph].
\bibitem{AF1}A. M. Frolov, Phys. Rev. A {\bf 57}, 2436 (1998).
\bibitem{VK2000}V. I. Korobov, Phys. Rev. A {\bf 61}, 064503 (2000).
\bibitem{KP2001}K. Pachucki, Phys. Rev. A {\bf 63}, 032508 (2001).
\bibitem{KM2008}A. A. Krutov and A. P. Martynenko, Phys. Rev. A {\bf 78}, 032513 (2008).
\bibitem{LM1}S. D. Lakdawala and P. J. Mohr, Phys. Rev. A {\bf 22}, 1572 (1980).
\bibitem{LM2}S. D. Lakdawala and P. J. Mohr, Phys. Rev. A {\bf 29}, 1047 (1984).
\bibitem{Borie}E. Borie, Z. Phys. A {\bf 291}, 107 (1979).
\bibitem{MT}P. J. Mohr, B. N. Taylor and D. B. Newell, Rev. Mod. Phys. {\bf 80}, 633 (2008).
\bibitem{t4}V. B. Berestetskii, E. M. Lifshits and L. P. Pitaevskii, Quantum
Electrodynamics, Nauka, Moscow, 1980.
\bibitem{M1}A. P. Martynenko, Phys. Rev. A {\bf 71}, 022506 (2005).
\bibitem{M2}A. P. Martynenko, JETP {\bf 101}, 1021 (2005);
hep-ph/0412250.
\bibitem{EM2009}E. N. Elekina, A. P. Martynenko, arXiv:0909.2759 [hep-ph].
\bibitem{Hameka}H. F. Hameka, Jour. Chem. Phys. {\bf 47}, 2728 (1967).
\bibitem{VAF}V. A. Fok, Principles of Quantum Mechanics, Nauka, Moscow, 1976.
\bibitem{GY}H. Grotch and D. R. Yennie, Rev. Mod. Phys.
{\bf 41}, 350 (1969).
\bibitem{BY}G. T. Bodwin and D. R. Yennie, Phys. Rev. D {\bf 37},
498 (1988).
\bibitem{M2000}R. N. Faustov and A. P. Martynenko, Phys. Atom. Nucl.
{\bf 65}, 265 (2002); Phys. Atom. Nucl. {\bf 63}, 845 (2000); Phys.
Rev. A {\bf 67}, 052506 (2003).
\bibitem{mcc}J. S. McCarthy, I. Sick and R. R. Whitney, Phys. Rev. C
{\bf 15}, 1396 (1977).
\bibitem{Drake}D. C. Morton, Q. Wu and G. W. F. Drake, Phys. Rev. C
{\bf 73}, 034502 (2006).
\bibitem{RA}R. Arnowitt, Phys. Rev. {\bf 92}, 1002 (1953).
\bibitem{BE}S. J. Brodsky and G. W. Erickson, Phys. Rev. {\bf 148}, 26 (1966).
\bibitem{KP}N. Kroll and F. Pollack, Phys. Rev. {\bf 84}, 594 (1951).
\bibitem{KKS}R. Karplus, A. Klein and J. Schwinger, Phys. Rev. {\bf 84}, 597 (1951).
\end{thebibliography}
\end{document}